\documentstyle[preprint,aps]{revtex}

\begin{document}
\title{Mixing of three neutrinos in matter}
\author{G. Cheng\thanks{%
Email: gcheng@ustc.edu.cn}}
\address{CCAST (World Laboratory), P.O. Box 8730, Beijing 100080, P.R. China%
\\
and\\
Astronomy, and Applied Physics Department, USTC., \\
Hefei, Anhui, 230026, P.R. China\thanks{%
Mailing address}}
\maketitle

\begin{abstract}
Explicit analytical expression is derived for mixing matrix of three
neutrinos in matter using a set of direct vacuum physical parameters.
Results are presented in simple, symmetrical form. The physical contents are
more clear than using of traditional mixing angles. There is no problem about 
mixing orders.
\end{abstract}

\pacs{14.60.Pq}

%%%%%%%%%%%%%%%%%%%%%%%%%main text%%%%%%%%%%%%%%%%%%%%%%%%%%%%%%%%%%%%

\newpage \baselineskip20pt

\section{Introduction}

The deficit of the solar neutrinos leads to the studies of neutrino
oscillations in matter \cite{davis}. Many important results have been
obtained for two flavors. An important fact, discovered more than ten years
ago, is that there is resonance of the mixing angle in matter due to the
interaction of charged current \cite{msw}. Because of the increase
of mixing parameters the situation is more complex in three flavors than two 
\cite{kuo}.

The supposition of small vacuum mixing angles between three flavor neutrinos
is accepted as a natural extension of small mixing angles between quarks in
the earlier period. Usually, one uses the mixing matrix written in a form of
three submatrix product, each containing a mixing angle \cite{kuo2}.
In recent years, the large vacuum mixing angle between $\nu _\mu $ and $\nu
_\tau $ has been established almost certainly by the atmosphere neutrino
experiments \cite{sk}. Large vacuum mixing angle between $\nu _e$
and $\nu _\mu $ has been suggested also by some authors in combined research
of solar neutrinos and the day night effect etc. 
\cite{bahcall}. Generally, the $\nu _e$ and $\nu _\tau $ mixing angle is
believed to be small \cite{chooz}. With small mixing angles, the mixing
order is not important. There are no difficulties in the explanation of
their physical meanings. However, with large mixing, the product order
become important. Therefor, it is possible that some important physical
characteristic is hidden uncovered when the order of the product is not
chosen in a good fashion. Some time it leads to misunderstanding.

A better general mathematical framework is desirable. We presented one in
this paper using the direct physical parameters $\eta _i$. In present
framework, the explicit analytical results are obtained. The mass
eigenvalues, eigenfunctions, and mixing parameters in matter are represented
analytically by vacuum mass, mixing parameters, and the effective potential $%
A$ from charged current interaction. They are expressed in completely
systematical form. There is no problem about mixing orders. And the physical
connotations are exposed more clearly. It will be valuable as a powerful
tool for further researches in three flavors neutrino oscillation in matter.
We restrict ourselves to a most useful but special case, the three flavor
mixing in matter with large mass differences, but without the small vacuum
mixing angle restriction. For simplicity, the $CP$ phase is ignored. There
is no difficulty to add it in this theory. We discuss the charged current
interaction between $\nu _e$ and $e$ only.

\section{Mixing matrixes between the eigenvectors of flavor and
masses\ in matter and flavor and mass in vacuum}

\ 

The mixing matrix $U^{m}$ which links the neutrino mass eigenstates in
matter to the neutrino flavor eigenstates is defined as 
\begin{equation}
\left\langle \nu _{\alpha }\right| =U_{\alpha u}^{m}\left\langle \nu
_{u}^{m}\right| 
\end{equation}%
where $\left| \nu _{\alpha }\right\rangle ,\;\alpha =e,\mu ,\tau ,\;$and $%
\left| \nu _{u}^{m}\right\rangle ,\;u=1,2,3,\;$represent the three
orthogonal and normalized eigenvectors of flavors and mass in matter of the
neutrino and $\left\langle \nu _{\alpha }\right| \;$and $\left\langle \nu
_{u}^{m}\right| \;$are their dual respectively. Summation convention for
repeated indices has been used. We always use this rule below otherwise we
indicate especially. Thus%
\begin{equation}
U_{\alpha u}^{m}=\left\langle \nu _{\alpha }\right| \nu _{u}^{m}\rangle 
\end{equation}%
If we use $\left| \nu \right\rangle \;$to represent the state vector of a
neutrino and $\nu _{\alpha }\;\alpha =e,\mu ,\tau $ and$\;\nu _{u}^{m}$ $%
u=1,2,3$ to express its components in the two bases spanned by $\left| \nu
_{\alpha }\right\rangle \;$and $\left| \nu _{u}^{m}\right\rangle $
respectively. That is 
\begin{equation}
\nu _{\alpha }=\left\langle \nu _{\alpha }\right| \nu \rangle
\end{equation}%
\begin{equation}
\nu_{u}^{m}=\left\langle \nu _{u}^{m}\right| \nu \rangle 
\end{equation}%
then, we have 
\begin{equation}
\nu _{\alpha }=U_{\alpha u}^{m}\nu _{u}^{m}
\end{equation}%
However because $\left\langle \nu _{u}^{m}\right| \nu _{\alpha }\rangle
=\left\langle \nu _{\alpha }\right| \nu _{u}^{m}\rangle $ is real in our
Hilbert space,\ so we have also 
\begin{equation}
\left| \nu _{\alpha }\right\rangle =U_{\alpha u}^{m}\left| \nu
_{u}^{m}\right\rangle 
\end{equation}%
In the same way, we define another transformation matrix U as%
\begin{equation}
\left\langle \nu _{\alpha }\right| =U_{\alpha i}\left\langle \nu _{i}\right| 
\end{equation}%
where $\left| \nu _{i}\right\rangle \;i=1,2,3\;$represent the three
orthogonal and normalized eigenvectors of mass in vacuum of the neutrino. We
have%
\begin{equation}
U_{\alpha i}=\left\langle \nu _{\alpha }\right| \nu _{i}\rangle 
\end{equation}%
Let 
\begin{equation}
\nu _{i}=\left\langle \nu _{i}\right| \nu \rangle 
\end{equation}%
We have also%
\begin{equation}
\nu _{\alpha }=U_{\alpha i}\nu _{i}
\end{equation}%
and 
\begin{equation}
\left| \nu _{\alpha }\right\rangle =U_{\alpha i}\left| \nu _{i}\right\rangle 
\end{equation}%
It should be noticed that we have used the $\alpha ,\beta ,...=e,\mu ,\tau ,$
the $u,v,...=1,2,3$ and the $i,j...=1,2,3$ as the indices of the neutrino
eigenvectors of flavor, mass in matter, and mass in vacuum respectively.

We would like to have a set of relations between the matrix elements $%
U_{\alpha u}^{m}$ and $U_{\alpha i},$ using the vacuum masses $m_{i}$ and
potential $A$ as parameters, and $U_{\alpha u}^{m}$ are expressed in
explicit analytical form. For this, we separate $U^{m}$ into two factors 
\begin{equation}
U^{m}=UW
\end{equation}%
Writing in component form 
\begin{equation}
U_{\alpha u}^{m}=U_{\alpha i}W_{iu}
\end{equation}%
it is easy to see W is a transformation matrix which connect the
eigenvectors of mass in vacuum and mass in matter. It can be expressed as 
\begin{equation}
W_{iu}=\left\langle \nu _{i}\right| \nu _{u}^{m}\rangle 
\end{equation}%
Now it is clear that our task is to search for the\ $\left| \nu
_{u}^{m}\right\rangle $ represented in the base spanned by eigenvectors of
mass in vacuum.

\section{Eigenvalues and eigenfunctoins of mass matrix in matter}

Writing in flavor representation, the mass matrix in matter of three
generation neutrinos is \cite{kuo2}:%
\begin{equation}
M_{f}^{2}=U\left( 
\begin{array}{lll}
m_{1}^{2} & 0 & 0 \\ 
0 & m_{2}^{2} & 0 \\ 
0 & 0 & m_{3}^{2}%
\end{array}%
\right) U^{\dagger }+\left( 
\begin{array}{lll}
A & 0 & 0 \\ 
0 & 0 & 0 \\ 
0 & 0 & 0%
\end{array}%
\right) 
\end{equation}%
Where $m_{i},$ $i=1,2,3$, is the neutrino vacuum masses. $A$ is an efficient
potential of the electron neutrino $\nu _{e}$ in matter. As usual, we
consider only charged current interaction between $\nu _{e}$ and the
electron $e$ in matter.%
\begin{equation}
A=2\sqrt{2}G_{f}N_{e}E
\end{equation}%
Where $G_{f}$\ is the Fermi constant, $N_{e}$ is the electron number density
of the matter, and $E$ is the energy of the neutrino.

Our aim is to search for the $W_{iu}=\left\langle \nu _{i}\right| \nu
_{u}^{m}\rangle .$ So it is necessary to write the matrix of mass in matter
in the vacuum mass representation%
\begin{equation}
.M_{m}^{2}=\left( 
\begin{array}{lll}
m_{1}^{2} & 0 & 0 \\ 
0 & m_{2}^{2} & 0 \\ 
0 & 0 & m_{3}^{2}%
\end{array}%
\right) +U^{\dagger }\left( 
\begin{array}{lll}
A & 0 & 0 \\ 
0 & 0 & 0 \\ 
0 & 0 & 0%
\end{array}%
\right) U  \label{m}
\end{equation}%
Let\ 
\begin{equation}
\eta =\left( 
\begin{array}{lll}
\eta _{1} & \eta _{2} & \eta _{3}%
\end{array}%
\right) =\left( 
\begin{array}{lll}
U_{e1} & U_{e2} & U_{e3}%
\end{array}%
\right) 
\end{equation}%
That is\ $\eta _{i}=U_{ei},$ $i=1,2,3$. It is easy to show%
\begin{equation}
U^{\dagger }\left( 
\begin{array}{lll}
A & 0 & 0 \\ 
0 & 0 & 0 \\ 
0 & 0 & 0%
\end{array}%
\right) U=A\eta ^{\dagger }\eta =A\left( 
\begin{array}{lll}
\eta _{11} & \eta _{12} & \eta _{13} \\ 
\eta _{21} & \eta _{22} & \eta _{23} \\ 
\eta _{31} & \eta _{32} & \eta _{33}%
\end{array}%
\right) 
\end{equation}%
where\ $\eta _{ij}=\eta _{i}\eta _{j}=\eta _{ji}.$\ We do not accept the
traditional mixing angles, but instead of the use of original physical
parameters $\eta _{i}=U_{\alpha i}=\left\langle \nu _{\alpha }\right| \nu
_{i}\rangle $ in our theoretical framework. By this, all the expressions are
simple, the results are expressed in symmetrical form, and the physical
contents are clear, not only for the mixing matrix discussed in this paper
but also for the expressions of resonance, propagation equation, and so on,
discussed in followed papers. There is a relation between three $\eta _{i}.$
It is easy to prove that%
\begin{equation}
\eta _{1}^{2}+\eta _{2}^{2}+\eta _{3}^{2}=1
\end{equation}%
Taking the trace to $\eta _{ij}$ matrix, we obtain%
\begin{equation}
{\rm Tr}\left( 
\begin{array}{lll}
\eta _{11} & \eta _{12} & \eta _{13} \\ 
\eta _{21} & \eta _{22} & \eta _{23} \\ 
\eta _{31} & \eta _{32} & \eta _{33}%
\end{array}%
\right) \equiv \sum\limits_{i=1}^{3}\eta _{i}^{2}={\rm Tr}\left[ U^{\dagger
}\left( 
\begin{array}{lll}
1 & 0 & 0 \\ 
0 & 0 & 0 \\ 
0 & 0 & 0%
\end{array}%
\right) U\right] =1
\end{equation}%
Thus, we have two independent parameters, not three. We can go further to
simplify the Eq.(\ref{m}) as usual by subtracting the same mass $%
(m_{1}^{2}+m_{2}^{2})/2$ from the $M_{m}^{2}.$ The mass matrix in matter can
be written as 
\begin{equation}
M_{m}^{2}=\frac{1}{2}\left( m_{2}^{2}+m_{1}^{2}\right) +\overline{M}_{m}^{2}
\end{equation}%
Where\ 
\begin{equation}
\overline{M}_{m}^{2}=\left( 
\begin{array}{lll}
-\Delta m_{1}^{2} & 0 & 0 \\ 
0 & \Delta m_{1}^{2} & 0 \\ 
0 & 0 & \Delta m_{2}^{2}%
\end{array}%
\right) +A\left( 
\begin{array}{lll}
\eta _{11} & \eta _{12} & \eta _{13} \\ 
\eta _{21} & \eta _{22} & \eta _{23} \\ 
\eta _{31} & \eta _{32} & \eta _{33}%
\end{array}%
\right) 
\end{equation}%
and%
\begin{equation}
\Delta m_{1}^{2}=\frac{1}{2}\left( m_{2}^{2}-m_{1}^{2}\right) ,\quad \Delta
m_{2}^{2}=m_{3}^{2}-\frac{1}{2}\left( m_{2}^{2}+m_{1}^{2}\right) 
\end{equation}%
The eigenvalue equation of $\overline{M}%
_{m}^{2}=M_{m}^{2}-(m_{1}^{2}+m_{2}^{2})/2$ becomes 
\begin{equation}
\overline{M}_{m}^{2}\left( 
\begin{tabular}{l}
$\nu _{1}^{m}$ \\ 
$\nu _{2}^{m}$ \\ 
$\nu _{3}^{m}$%
\end{tabular}%
\ \right) =\lambda \left( 
\begin{tabular}{l}
$\nu _{1}^{m}$ \\ 
$\nu _{2}^{m}$ \\ 
$\nu _{3}^{m}$%
\end{tabular}%
\ \right) 
\end{equation}%
Then, the dependence on the three vacuum masses $m_{i}^{2},$ $(i=1,2,3,)$ is
reduced into two $\Delta m_{i}^{2},$ $(i=1,2)$. The condition for a
nontrivial solution to be existence for the eigenstate equation is 
\begin{equation}
\det \left( 
\begin{array}{ccc}
-\Delta m_{1}^{2}+A\eta _{11}-\lambda  & A\eta _{12} & A\eta _{13} \\ 
A\eta _{21} & \Delta m_{1}^{2}+A\eta _{22}-\lambda  & A\eta _{23} \\ 
A\eta _{31} & A\eta _{32} & \Delta m_{2}^{2}+A\eta _{33}-\lambda 
\end{array}%
\right) =0
\end{equation}%
Expanding it, we get a algebraic equation in third order. 
\begin{equation}
\lambda ^{3}+a\lambda ^{2}+b\lambda +c=0
\end{equation}%
where 
\begin{equation}
a=-(A+\Delta m_{2}^{2})
\end{equation}%
\begin{equation}
b=-(\Delta m_{1}^{2})^{2}+\left[ \Delta m_{1}^{2}\left( \eta _{11}-\eta
_{22}\right) +\Delta m_{2}^{2}\left( \eta _{11}+\eta _{22}\right) \right] A
\end{equation}%
and 
\begin{equation}
c=(\Delta m_{1}^{2})^{2}\Delta m_{2}^{2}-\Delta m_{1}^{2}\left[ \Delta
m_{2}^{2}\left( \eta _{11}-\eta _{22}\right) -\Delta m_{1}^{2}\eta _{33}%
\right] A
\end{equation}%
The solutions of $\lambda $ are standard in algebra but somewhat
complicated. For completeness we write them at below but will use it seldom.
The three eigenvalues of mass in matter, $M_{u}^{2}$ $(u=1,2,3,),$ of the
three generation neutrinos are: 
\begin{equation}
M_{m,u}^{2}=\lambda _{u}+\frac{1}{2}\left( m_{1}^{2}+m_{2}^{2}\right) \quad
\quad \;u=1,2,3,
\end{equation}%
where 
\begin{equation}
\lambda _{u}=Y_{u}-\frac{a}{3},\qquad \;u=1,2,3,
\end{equation}%
\begin{equation}
Y_{1}=\beta _{1}+\beta _{2},\quad Y_{2}=\alpha _{1}\beta _{1}+\alpha
_{2}\beta _{2},\quad and\quad Y_{3}=\alpha _{1}\beta _{2}+\alpha _{2}\beta
_{1}
\end{equation}%
\begin{equation}
\alpha _{1}=-\frac{1}{2}+i\frac{\sqrt{3}}{2},\quad \alpha _{2}=-\frac{1}{2}-i%
\frac{\sqrt{3}}{2},
\end{equation}%
\begin{equation}
\beta _{1}=\left( \sqrt{Q}-\frac{q}{2}\right) ^{\frac{1}{3}},\quad and\quad
\beta _{2}=\left( -\sqrt{Q}-\frac{q}{2}\right) ^{\frac{1}{3}}
\end{equation}%
\begin{equation}
Q=\left( \frac{p}{3}\right) ^{3}+\left( \frac{q}{2}\right) ^{2}
\end{equation}%
\begin{equation}
p=b-\frac{1}{3}a^{2},\quad q=\frac{2}{27}a^{3}-\frac{1}{3}ab+c
\end{equation}

Having the eigenvalues solved, the eigenfunctions of mass in matter,
represented in the frame spanned by the vacuum mass eigenvectors, can be
obtained by solving the eigenstate equation, Eq.(25). Rewriting Eq.(25) in a
form as follows%
\begin{equation}
\begin{array}{l}
A\left( 
\begin{array}{ccc}
\eta _{11} & \eta _{12} & \eta _{13} \\ 
\eta _{21} & \eta _{22} & \eta _{23} \\ 
\eta _{31} & \eta _{32} & \eta _{33}%
\end{array}%
\right) \left( 
\begin{tabular}{c}
$\nu _{1}^{u}$ \\ 
$\nu _{2}^{u}$ \\ 
$\nu _{3}^{u}$%
\end{tabular}%
\ \right) 
=\left( 
\begin{array}{ccc}
\lambda _{u}+\Delta m_{1}^{2} & 0 & 0 \\ 
0 & \lambda _{u}-\Delta m_{1}^{2} & 0 \\ 
0 & 0 & \lambda _{u}-\Delta m_{2}^{2}%
\end{array}%
\right) \left( 
\begin{tabular}{c}
$\nu _{1}^{u}$ \\ 
$\nu _{2}^{u}$ \\ 
$\nu _{3}^{u}$%
\end{tabular}%
\ \right) 
\end{array}%
\end{equation}%
After routine algebra, we can obtain the solution. 
\begin{equation}
\nu _{i}^{m,u}=\langle \nu _{i}\mid \nu _{u}^{m}\rangle =N^{u}\frac{\eta _{i}%
}{\lambda _{u}-\Delta _{i}}\qquad u=1,2,3,\quad i=1,2,3
\end{equation}%
where 
\begin{equation}
\Delta _{1}=-\Delta m_{1}^{2},\quad \Delta _{2}=\Delta m_{1}^{2},\quad
\Delta _{3}=\Delta m_{2}^{2}\qquad i=1,2,3
\end{equation}%
and $\mid \nu ^{m,u}\rangle $ is the eigenvector corresponding the mass $%
M_{m,u}$, $u=1,2,3$, and $N^{u}$ is a normalized constant. It is given by 
\begin{equation}
N^{u}=\left[ \sum\limits_{i=1}^{3}\left( \frac{\eta _{i}}{\lambda
_{u}-\Delta _{i}}\right) ^{2}\right] ^{-\frac{1}{2}}\qquad u=1,2,3
\end{equation}%
In the above three equations, the repeated index $i$ does not subject to the
summing rule. Because $\lambda _{u}-\Delta _{i}=M_{m,u}^{2}-m_{i}^{2},$
Eq.(39) and Eq.(41) can also be written in a form using original parameters $%
M_{m,u}^{2}$ and $m_{i}.$%
\begin{equation}
\nu _{i}^{m,u}=\langle \nu _{i}\mid \nu _{u}^{m}\rangle =N^{u}\frac{\eta _{i}%
}{M_{m,u}^{2}-m_{i}^{2}}\qquad u=1,2,3,\quad i=1,2,3
\end{equation}%
\begin{equation}
N^{u}=\left[ \sum\limits_{i=1}^{3}\left( \frac{\eta _{i}}{%
M_{m,u}^{2}-m_{i}^{2}}\right) ^{2}\right] ^{-\frac{1}{2}}\qquad u=1,2,3,
\end{equation}%
However, writing in this form, we need three vacuum mass $m_{i}$, but in the
Eq.(39) and Eq.(41), it is only necessary to have two vacuum mass
differences known.

\section{The expression of mixing matrix in matter and the problem of mixing
order}

By Eq.(13) and Eq.(14), we have $U_{\alpha u}^{m}=U_{\alpha i}W_{iu}$ and $%
W_{iu}=\left\langle \nu _{i}\right| \nu _{u}^{m}\rangle =N^{u}\eta
_{i}/\left( \lambda _{u}-\Delta _{i}\right) =N^{u}\eta _{i}/\left(
M_{u}^{2}-m_{i}^{2}\right) .$ Then we have%
\begin{equation}
U_{\alpha u}^{m}=N^{u}\sum\limits_{i=1}^{3}U_{\alpha i}\frac{\eta _{i}}{%
\lambda _{u}-\Delta _{i}}=N^{u}\sum\limits_{i=1}^{3}U_{\alpha i}\frac{\eta
_{i}}{M_{u}^{2}-m_{i}^{2}}\quad \alpha =e,\mu ,\tau ,\;u=1,2,3
\end{equation}%
In particular, we have the very simple form for\ $\nu _{e}$\ mixing%
\begin{equation}
U_{eu}^{m}=N^{u}\sum\limits_{i=1}^{3}\frac{\eta _{i}^{2}}{\lambda
_{u}-\Delta _{i}}=N^{u}\sum\limits_{i=1}^{3}\frac{\eta _{i}^{2}}{%
M_{u}^{2}-m_{i}^{2}}\qquad u=1,2,3
\end{equation}

In the traditional theory, one accepts the form of a matrix with three
mixing angles \cite{kuo2}%
\begin{equation}
U=U_{2}U_{3}U_{1}
\end{equation}%
where 
\begin{equation}
U_{1}=\left( 
\begin{array}{ccc}
c_{1} & s_{1} & 0 \\ 
-s_{1} & c_{1} & 0 \\ 
0 & 0 & 1%
\end{array}%
\right) 
\end{equation}%
\begin{equation}
U_{2}=\left( 
\begin{array}{ccc}
1 & 0 & 0 \\ 
0 & c_{2} & s_{2} \\ 
0 & -s_{2} & c_{2}%
\end{array}%
\right) 
\end{equation}%
\begin{equation}
U_{3}=\left( 
\begin{array}{ccc}
c_{3} & 0 & s_{3} \\ 
0 & 1 & 0 \\ 
-s_{3} & 0 & c_{3}%
\end{array}%
\right) 
\end{equation}%
in which $c_{i}=\cos \theta _{i}$ and $s_{i}=\sin \theta _{i},\;i=1,2,3$. $%
\theta _{i}$ are called vacuum mixing angles. When all of the $\theta _{i}$
are small, the orders of the three rotations are not important. From
mathematical and physical view, expressions in different orders are
equivalent when the terms with higher order small quantities can be
neglected. In physics, they represent three successive rotation angles in
three different planes respectively. When any of the mixing angle $\theta
_{i}$ is large, we can not explain as that, at least to some of the angles.
In different orders, each $\theta _{i}$ represents a different physical
content. We call it the problem of orders. In the same way, mixing in matter
is also expressed as a product of three matrix in traditional theory 
\begin{equation}
U^{m}=U_{2}^{m}U_{3}^{m}U_{1}^{m}
\end{equation}%
where $U_{i}^{m}$ can be obtained using Eq.(47)-Eq.(49) by replacing the $%
c_{i}$ and $s_{i}$ by $c_{i}^{m}$ and $s_{i}^{m}$ respectively and let $%
c_{i}^{m}=\cos \theta _{i}^{m}$ and $s_{i}^{m}=\sin \theta
_{i}^{m},\;i=1,2,3.$ The $\theta _{i}^{m}$ is called mixing angles in
matter. The problem of mixing order is the same as that in vacuum mixing
case.

However, in our expressions, there is no order problem at all. The order
problem is more subtle in the resonance phenomena in matter. We will discuss
it in more details in a followed paper. In there, the advantage of our
present theoretical framework will be more remarkable.

\section{Conclusion}

We have given a set of explicit analytical expressions connecting mixing
matrix of three neutrinos in matter and vacuum parameters. In these formulae
we have used a set of new physical parameters directly, instead of those
used in traditional theory. Our results are simple, symmetrical, and
physically clear. Moreover, there is no order problem here.

\acknowledgements

I would like to express my sincere thanks to Prof. A. S. Hirsch, Prof. T. K.
Kuo and the members of High Energy Theory Group, Physics Department, Purdue
University where part of the work was completed, for the kind hospitality
extended to me during my visit from January to April, 2000. Many fruitful\
discussions with Prof. T. K. Kuo are very helpful to this work. I would like
to think Mr. Hai-Jun Pan and Mr. Tao Tu for their help in doing this work.
The author is supported in part by the National Science Foundation, China.

%\addtolength{\baselineskip}{-.3\baselineskip}

\end{document}